\documentclass[sigconf,screen,nonacm]{acmart}

\AtBeginDocument{%
  \providecommand\BibTeX{{%
    \normalfont B\kern-0.5em{\scshape i\kern-0.25em b}\kern-0.8em\TeX}}}

\begin{document}

\title[SynthScribe]{SynthScribe: Deep Multimodal Tools for Synthesizer Sound Retrieval and Exploration}

\author{Stephen Brade}
\affiliation{%
  \institution{University of Toronto}
  \city{Toronto}
  \country{Canada}
}
\email{stephen.brade@mail.utoronto.ca }

\author{Bryan Wang}
\affiliation{%
  \institution{University of Toronto}
  \city{Toronto}
  \country{Canada}
}
\email{bryanw@dgp.toronto.edu}

\author{Mauricio Sousa}
\affiliation{%
  \institution{University of Toronto}
  \city{Toronto}
  \country{Canada}
}
\email{mauricio@dgp.toronto.edu}

\author{Gregory Lee Newsome}
\affiliation{%
  \institution{University of Toronto}
  \city{Toronto}
  \country{Canada}
}
\email{greg.newsome@utoronto.ca}

\author{Sageev Oore}
\affiliation{%
  \institution{Dalhousie University}
  \city{Halifax}
  \country{Canada}
}
\email{sageev@dal.ca}

\author{Tovi Grossman}
\affiliation{%
  \institution{University of Toronto}
  \city{Toronto}
  \country{Canada}
}
\email{tovi@dgp.toronto.edu}

\renewcommand{\shortauthors}{Brade et al.}

\begin{abstract}
    Synthesizers are powerful tools that allow musicians to create dynamic and original sounds. Existing commercial interfaces for synthesizers typically require musicians to interact with complex low-level parameters or to manage large libraries of premade sounds. To address these challenges, we implement SynthScribe --- a fullstack system that uses multimodal deep learning to let users express their intentions at a much higher level. We implement features which address a number of difficulties, namely 1) searching through existing sounds, 2) creating completely new sounds, and 3) making meaningful modifications to a given sound. This is achieved with three main features: a multimodal search engine for a large library of synthesizer sounds; a user centered genetic algorithm by which completely new sounds can be created and selected given the users preferences; a sound editing support feature which highlights and gives examples for key control parameters with respect to a text or audio based query.  The results of our user studies show SynthScribe is capable of reliably retrieving and modifying sounds while also affording the ability to create completely new sounds that expand a musicians creative horizon. 
\end{abstract}

 \begin{teaserfigure}
   \centering
   \includegraphics[width=\textwidth]{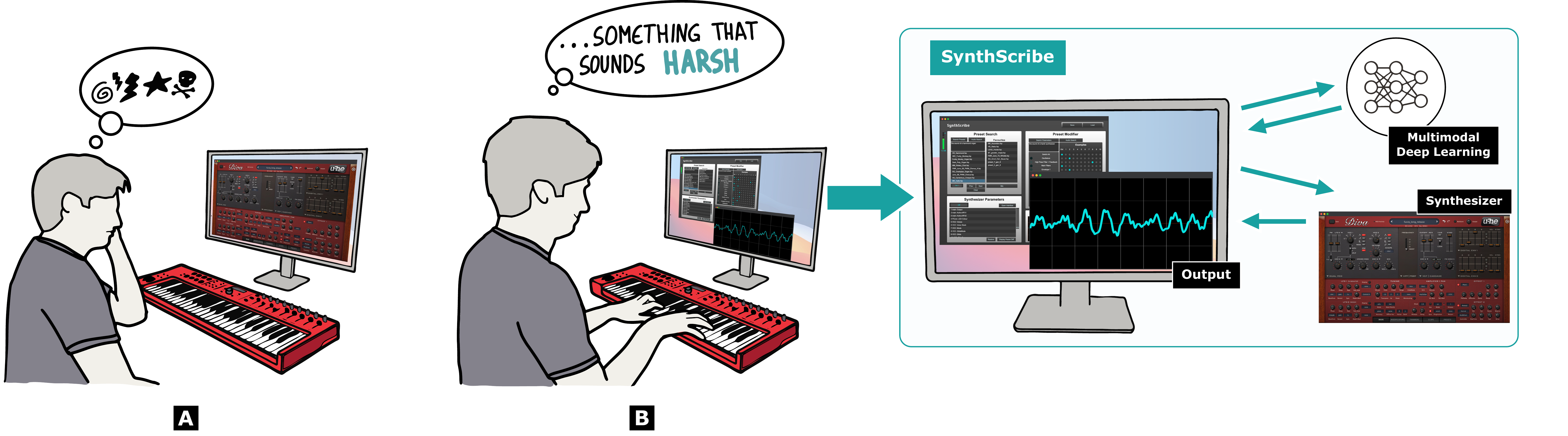}
   \caption{(A) a scene depicting a typical interaction with a synthesizer in which a musician has to manage hundreds of low-level parameters to get a desirable sound. (B) A contrasting interaction in which the musician uses SynthScribe to express their desires at a much higher level. This is achieved by using multimodal deep learning in the backend to help a user communicate their desires to a synthesizer, leading to a much better result. }
   \Description{A comic strip depicting a musicians experience using a synthesizer without SynthScribe and their experience using a synthesizer with SynthScribe. Without SynthScribe, the user is frustrated with the complicated interface for the synthesizer and is unable to create the harsh sound that they desire. With SynthScribe, they are considerably more satisfied and are able to create the harsh sound they are looking for.}
   \label{fig:teaser}
\end{teaserfigure}

\maketitle

\section{Introduction}

Synthesizers enable musicians to create rich and complex timbres — unique sounds that can be played on a keyboard or through other digital interfaces that allow composers to expand their timbral vocabulary beyond a typical set of instruments (e.g. guitars, violin, timpani, etc.). To use a synthesizer, musicians must learn to manage hundreds of control parameters that modify the various aspects of a sound (Fig.~\ref{fig:teaser}A). Many musicians typically rely on large banks of preset sounds which are designed by skilled engineers and packaged with a synthesizer as starting points for musicians. Allegedly, 9 out of 10 Yamaha DX7s (a popular synthesizer) that are brought in for servicing are returned with their default presets still intact, suggesting that default sounds are heavily relied on and that many musicians rarely succeed in creating unique sounds of their own~\cite{seago_critical_2004}. This may partly be due to the high dimensionality of synthesizer controls and their non-intuitive naming conventions. Synthesis parameters are often named after quantifiable aspects of sounds which are disconnected from typical experiential descriptions of sounds~\cite{seago_critical_2004}. 

Music technology companies and researchers have developed commercial tools and experimental systems that help users manage preset sounds and low-level synthesis parameters. Analog Lab V by Arturia~\footnote{Arturia's Analog Lab V: \url{https://www.arturia.com/products/software-instruments/analoglab-v/overview}} provides a bank of 5840 preset sounds which are labelled with various semantic attributes which can be accessed by the user via their keyword search engine. Music technology companies may also design macrocontrols for their synthesizers --- more intuitive controls which combine multiple control parameters into one. However, presets are often reused by multiple musicians which diminishes originality and macrocontrols are not customizable and limit the expressivity of the synthesizer as they project the set of possible sounds into a more limited, lower dimensional space. In response to this, machine learning and intelligent interface research has attempted to automatically create customizable and intuitive macrocontrols through active learning or by using latent representations of synthesizer sounds to craft macrocontrols~\cite{esling_flowsynth_2020, esling_universal_2019, huang_active_2014}. Researchers have also developed sound matching algorithms that would let users find the synthesis parameters that best replicate any piece of recorded audio \cite{horner1993machine, lai2006automated, masuda2021synthesizer}.

However, these prior techniques either require users to laboriously label sounds or for a new model to be trained per synthesizer. Other techniques like sound matching rely on the user to supply examples of desirable sounds, limiting the user to sounds that exist in prior music or in their environment. In this work, we combine several features in a novel system that can help users search for, modify, and create completely new synthesizer sounds by using text and audio concurrently as an intuitive control modality without training a new model or requiring user annotations.

To this end, we present SynthScribe (Fig.~\ref{fig:teaser}B) --- a full stack system that leverages multimodal deep learning to help musicians work with an existing synthesizer. Guided by our formative interviews with musicians, we implement three features which leverage LAION-CLAP \cite{laionclap2023} --- a multimodal deep learning model pretrained on music and general audio. We first implement a multimodal search that allows users to navigate through a bank of synthesizer sounds using either text or existing synthesizer sounds. We also create a user-centered genetic algorithm where hundreds of new synthesizer sounds can be created by mixing together a list of a musician's favourite presets; these new sounds are then recorded and embedded with LAION-CLAP on the fly, allowing users to rapidly discover completely new soundscapes through the multimodal search feature. Once a user has found a sound that nearly meets their desires, they can make meaningful edits to it using the preset modification feature which highlights important groups of parameters with respect to a text or audio query and provides examples of how to modify those parameters to achieve a desired effect. 

To evaluate this approach, we performed two user studies. In the first study, we evaluate the ability of LAION-CLAP to search through synthesizer sounds and our preset modification feature's effectivness at providing meaningful modifications. The results of this study showed that LAION-CLAP provides a reliable foundation for our search feature and that our preset modification feature is capable of making meaningful modifications to a sound. In the second study, we complete a series of free usage observations and find evidence that our system can be used to save time for both professional and amateur musicians alike in addition to helping musicians find pleasant yet surprising sounds that inspire creativity through our genetic mixing feature.

In summary, we make the following contributions:
    \begin{itemize}
        \item A novel system, SynthScribe, that uses multimodal deep learning to enable the use of text and audio for the intuitive control of a synthesizer.
        \item The results of a user study evaluating our systems ability to retrieve and modify existing synthesizer sounds; thus, evaluating LAION-CLAP's effectiveness as a foundation for these tasks.
        \item The results of free usage observations with musicians demonstrating that our system assists musicians in efficiently finding and creating new sounds while also providing them with unexpected sounds that inspire new creative directions.
    \end{itemize}

\section{Related Work}

\subsection{Multimodal Deep Learning}
Multimodal deep learning aims to establish relationships between modalities of data that arise concurrently. In recent years, contrastive learning approaches have proven to be useful in creating joint representations of data modalities that can be used for multimodal search or generative tasks. In the image domain, CLIP~\cite{clip} has been used as an engine for multimodal search~\cite{clip-retrieval} or as a key component in generative models like DALL-E 2~\cite{dalle} and Stable Diffusion~\cite{sd}. In audio, there is a breadth of research on deep learning models like LAION-CLAP \cite{laionclap2023} that create joint representations of text and audio, often with a focus on music ~\cite{manco_multimodal_2023, huang2022mulan}. Some preliminary work has created similar representations specifically for musical timbres and text~\cite{jonason2022timbreclip}. These models are potentially useful for musicians but must be embedded into a well designed system if they are to be integrated into a musicians workflow. In this work, we bridge this gap by experimenting with several novel features by recording and embedding synthesizer sounds using LAION-CLAP~\cite{laionclap2023}. By leveraging the ability to connect text and audio, we can help musicians explore existing sounds and create thousands of new ones while also helping them meaningfully modify any sound.

\subsection{Assisted Sound Synthesis}
A significant body of research in music technology has been devoted to facilitating sound synthesis by experimenting with novel interface design and/or the application of existing machine-learning techniques. In interface design, researchers have focused on designing expressive methods of control for synthesizers by representing the parameter space in touch-based interfaces~\cite{tc_11, pseudo_intention} or by allowing for direct manipulation by representing waveforms in a pin-based shape-shifting display or by letting users sketch the shapes of sound waves~\cite{colter_soundforms_2016, nakanishi_nakanisynth_2016}. Early work in machine learning focused on the application of evolutionary algorithms to optimize synthesizer parameters to match a given sound~\cite{lai2006automated, horner1993machine}. Sound matching has also recently been attempted via deep learning through the use of differentiable digital signal processing \cite{masuda2021synthesizer}. Other developments have focused on creating intuitive macrocontrols via active learning of semantic qualities of synthesizer sounds \cite{huang_active_2014} or by training latent representations of synthesis parameters using normalizing flows which capture high-level perceptual relationships between sounds \cite{esling_flowsynth_2020, esling_universal_2019}. We build on these techniques by providing musicians with an intuitive control modality for synthesizers without requiring a new model to be trained or the laborious collection of user annotations. We achieve this in SynthScribe by implementing several novel features on top of an existing multimodal deep learning model to facilitate the use of synthesizers.

\subsection{HCI and Parameter Exploration Problems}
There is a significant breadth of work in HCI which has focused on strategies for exploring design spaces that are governed by control parameters. Geppetto \cite{geppetto}, for example, implemented a system which helped users generate expressive movements in robots by using a data driven approach to connect complex control parameters with desired semantic behaviours. An earlier work, Attribit \cite{attribit}, allowed users to create visual content associated with a given semantic attribute, abstracting the design process away from parameters. Other work has focused on implementing intuitive control spaces for image editing and 3D design \cite{editing_outdoor_scenes, semantic_shape_editing_deformation_handles, crowd_powered_visual_design_exp, dream_lens}, animation \cite{optimo}, or the appearance of material textures \cite{material_appearance}.  In another vein, BO as Assistant \cite{koyama2022bo} retools Bayesian Optimization (BO) to let designers freely explore a parameter space while calculating promising regions for exploration on the fly. Other works have used BO to help users manage control parameters in animation \cite{animation}. Like the tasks these approaches were designed for, sound synthesis requires users to optimize a set of control parameters to match their creative intention. With SynthScribe, we facilitate interaction with low level control parameters by using multimodal deep learning as a foundation for searching through sounds and for providing examples of parameter adjustments that achieve a desired effect.

\section{Formative Interviews}
To understand current practices and identify key difficulties experienced by synthesizer users, we interviewed eight working musicians who use synthesizers to enhance their creative practice. The interviews were conducted via half-hour-long video conference calls, and the participants were compensated with 20 CAD. The musicians had diverse backgrounds with some using their expertise to create music for films or podcasts and others who collaborate with other musicians as producers. Although all were working musicians, each had varying comfort levels when working with synthesizers. In total, we interviewed three musicians who described themselves as experts in sound synthesis (P1, P2, P3), three intermediates (P5, P7, P8), and two novices (P4, P6). Each musician was asked questions about their current strategies and also their approach when they first started using synthesizers.

\subsubsection{The Trouble of Finding the Right Sound}
All participants discuss their varying reliance on preset sounds and their strategies for finding a fitting preset. Novices typically rely on presets due to the overwhelming nature of synthesizer parameters. P1 describes their first experience with a synthesizer interface as "sensory overload" while P4 explains that "earlier on, I'd be way too intimidated to touch things." Jargon also contributes to these difficulties. P2 comments while showing a synthesizer interface using screen share that "there are just so many words here that I just didn't understand going into it until I really studied synthesis." Musicians that are skilled with synthesizers tend to use preset sounds when they are under time pressure. P1 describes how they will use presets when submitting a score for a project that has a short time horizon.

To search through presets, some participants like P1 mention using software packages like Arturia's Analog Lab V which contain rich natural language labels for their 5840 presets which are utilized by a keyword search feature. These text labels are not always available for all synthesizers as other participants (P3, P4, P7, P8) mention taking a brute force approach and scrolling through large lists of presets to find one that fits with a musical context. Often large banks of preset sounds are organized into folders which pertain to certain types of synthesizer sounds (Leads, Pads, Bass, etc.). Beyond that, presets are named but as P1 notes these names are "hit or miss". Other participants (P3, P5, P8)  express a similar sentiment with P6 and P7 stating that they largely ignore preset names before they have listened to their sound. P1, P2, and P8 explain that specific names that reference a recognizable synthesizer sound can be useful. For example, P1 describes a preset named "Beat It" which they describe as an instantly recognizable sound to people familiar with that Michael Jackson song.

\subsubsection{The Trouble of Modifying Sounds}
Presets are not always satisfactory, however, and often need to be modified to be useful. P4 notes that presets "sound good in isolation and usually don't sound good within the context of the work". All other participants express similar sentiments, emphasizing that they prefer to modify preset sounds or design a completely new sound from scratch when given the time. Other participants (P1, P2, P6, P7, P8) describe how part of their collaboration with other musicians or stakeholders involves receiving feedback and incorporating feedback on synthesizer sounds. For example, P1 describes a situation in which they were requested by a singer to make the synthesizer playing the harmony to sound "more dreamy". 

To modify a sound as a novice, some participants (P3, P5, P7) relied on random exploration of synthesizer parameters while others like P4 and P6 would opt to just use presets due to being intimidated by synthesizer interfaces. As professionals, the ability to modify sounds comes with a detailed understanding of the parameters that allows them to comprehend the relationship between synthesis parameters and human perception. To develop this skill, a useful app mentioned by three participants (P2, P3, P6) is Syntorial ~\footnote{Syntorial: \url{https://www.syntorial.com/}}, a tool that helps musicians bring the timbres they hear in their head to life. This tool breaks synthesizers down into their fundamental components --- groups of parameters which affect specific aspects of a sound --- and then uses gamification to train musicians to connect those parameters with what they hear. Although useful, Syntorial is a learning tool and does not provide features that immediately facilitate the use of synthesizers

\subsubsection{The Importance of Originality}
Many participants (P1, P3, P4, P7) express the importance of originality in their artistic pursuits. P7 uses synthesizers because they remove limitations on their music noting that "if you get an acoustic guitar, you're kind of limited to how many genres or styles that you can make because you have the same timbre of an acoustic guitar."  P1 mentions that in film music "people are always trying to find the next new thing, and find new unique sounds". Conversely, P3 and P4 describe that synthesizer music is not noteworthy without a unique artistic signature. P3 describes music that was demoed in their classes, explaining that it began to sound redundant when the same presets are used by multiple students saying that "everyone's demo reels sound the same to a certain point. It's like, I know that preset." P4 emphasizes the importance of incorporating their own synthesizer sounds in the music they produce, expressing that "if you become too generic of a producer, then you don't have your own voice". These experiences point to the value that musicians place on finding unique sounds as it is a part of their strategy for cultivating a distinct artistic identity.
\section{Design Goals}
Using our formative interviews as an inspiration, we developed three goals to guide the design and development of SynthScribe to facilitate the use of synthesizers.

\vspace{6pt}
\noindent
\textbf{D1. Assisting Users in Finding Preset Sounds.} 
  Participants mention having access to large preset libraries without a way to meaningfully search through presets. Since presets provide a useful foundation for many musicians, we intend to design a system that can search through presets irrespective of whether those presets come with rich text labels.

\vspace{6pt}

\noindent \textbf{D2. Facilitating  Finetuning of Existing Presets.} 
SynthScribe should also enhance a musician's ability to meaningfully change preset sounds, enabling users to adapt a synth sound to their needs or to rapidly incorporate feedback from other stakeholders. 

\vspace{6pt}

\noindent \textbf{D3. Supporting the Creation of New Sounds.} 
SynthScribe should also enable users to create new sounds that go beyond presets, driving even novice users towards originality.
\section{SynthScribe}

\begin{figure*}[t]
  \centering
  \includegraphics[width=1\textwidth]{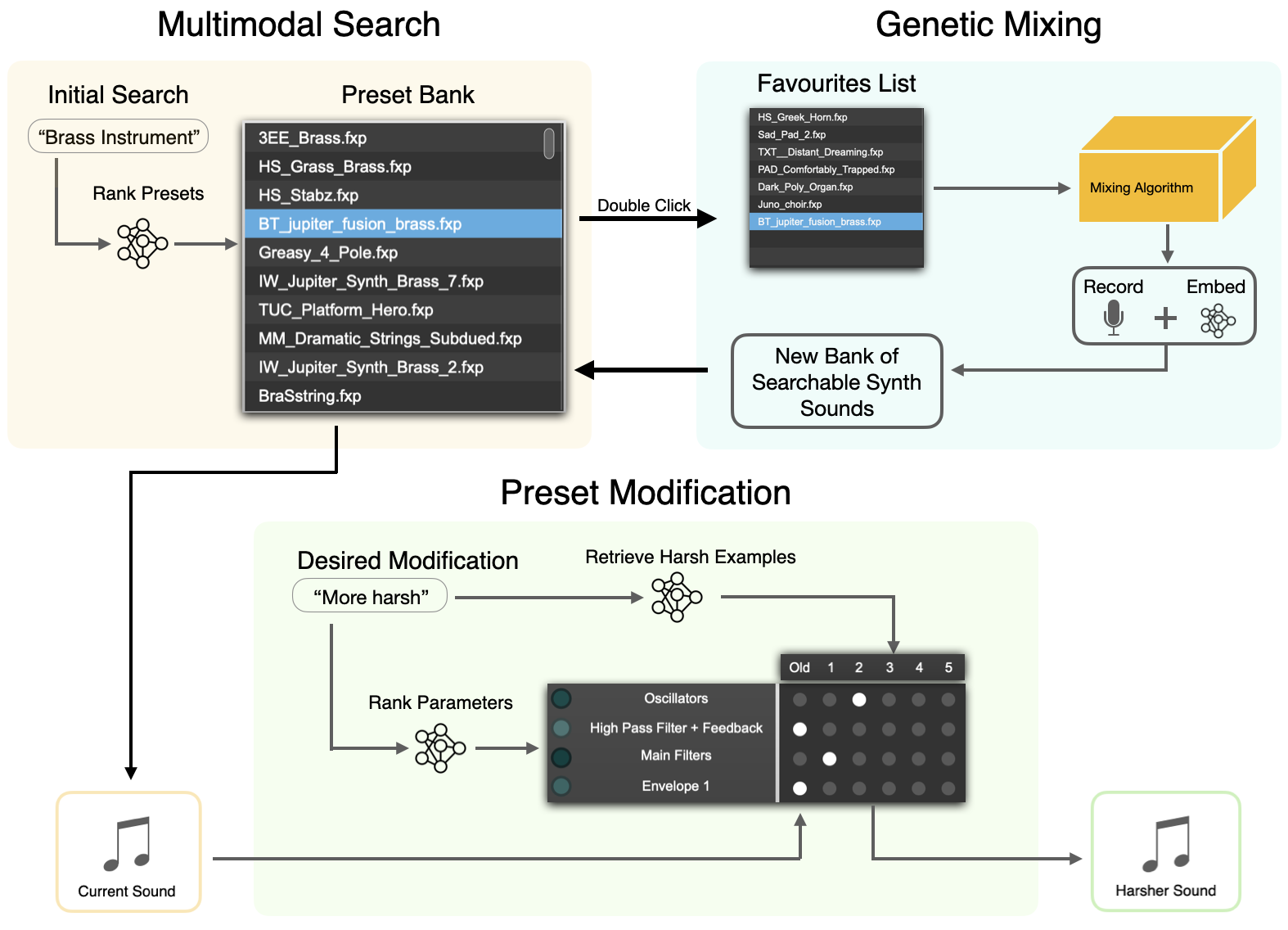}
  \caption{Figure displaying workflow that SynthScribe implements. Starting in the top left (orange), a user can use text to retrieve relevant synthesizer sounds from a default preset bank. To create completely new sounds, they can generate a completely new preset bank by using Genetic Mixing (turquoise) to create unique combinations of their favourite sounds from the default preset bank. Given a sound that is close to their desires, they can send it to the Preset Modification feature (green). Users can then express their desires for the modification in text and are provided with examples of how they can change the parameters to achieve a better sound.}
  \Description{Flowchart depicting the SynthScribe workflow. The first cell depicts the multimodal search feature which uses LAION-CLAP to search through lists of synthesizer sounds. This cell is connected to the genetic mixing cell which shows how synthesizer sounds can be selected as favourites and then mixed to create completely new sounds. The multimodal search cell is also connected to the Preset Modification Cell which takes in a synthesizer sound and modifies it with respect to a text query (e.g. "more harsh").}
  \label{fig:workflow}
\end{figure*}

SynthScribe builds several features on top of the Diva Synthesizer by U-He~\footnote{Diva Synthesizer by U-He: \url{https://u-he.com/products/diva/}} which creates a workflow displayed in Fig.~\ref{fig:workflow}. First, musicians can use a flexible multimodal search where they can use text or other presets to search through a bank of 3529 preset sounds. Users can expand this list of presets by choosing to randomly mix the parameters of a group of selected presets via the Genetic Mixing feature. This will create a new generation of synthesizer sounds that can be queried with the same multimodal search. Next, users can modify any synthesizer sound by using the Preset Modification feature. This feature lets users execute a parameter search highlighting which groups of parameters are important with respect to a text query that describes a desired effect and can make changes to their initial sounds by providing ten example changes per parameter group. At any time users can see a list of parameters or the default Diva interface. When using the preset modification feature, the parameter list can be set to show only parameters that have been changed with respect to a user's starting sound. This lets them narrow the control space to only crucial parameters when they discover a change moving toward their desired effect.

\begin{figure*}[t]
  \centering
  \includegraphics[width=1\textwidth]{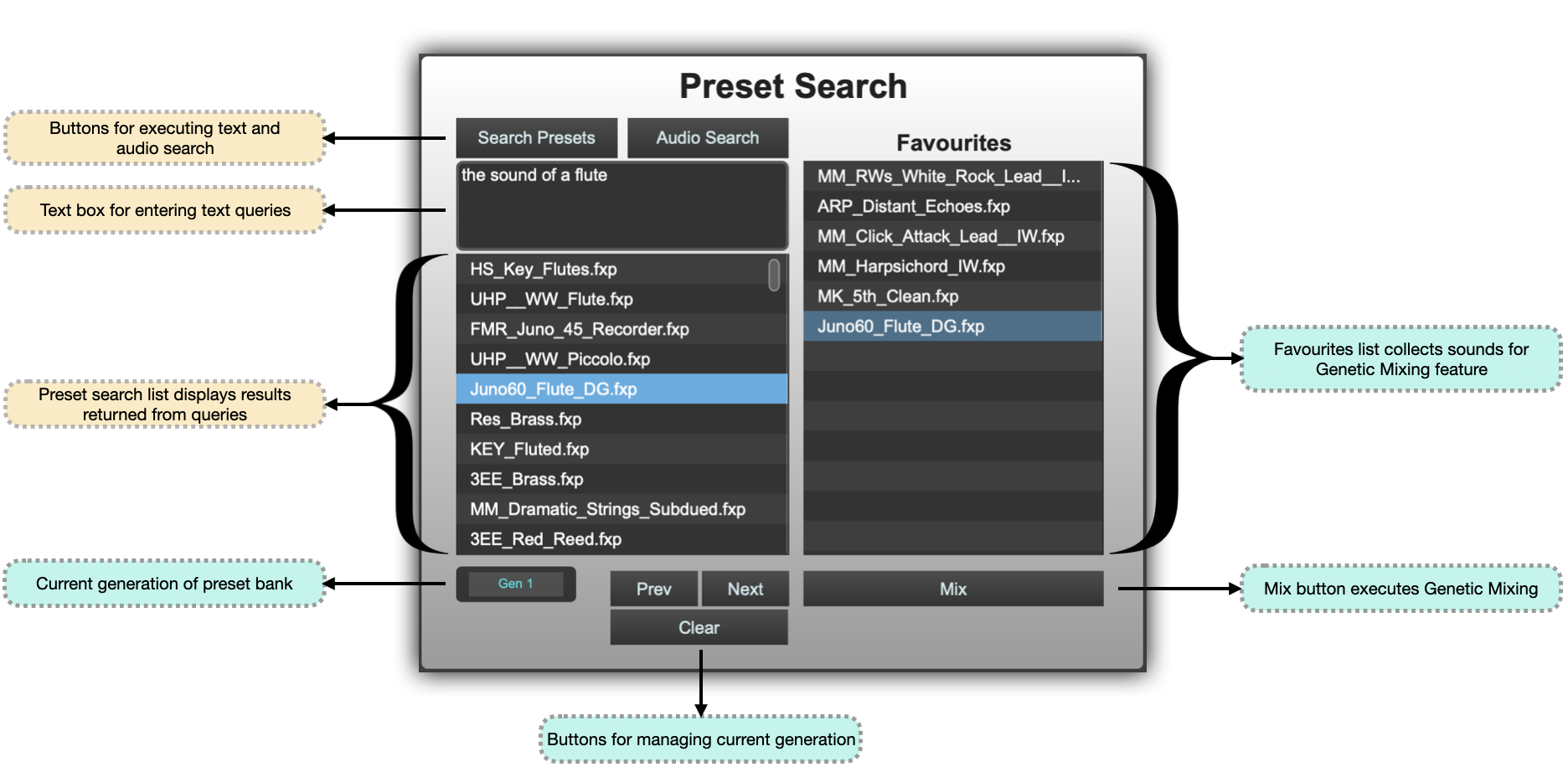}
  \caption{Interface for the window where users can search through the default preset bank (visualized in this figure) or presets generated with the Genetic Mixing feature. Key functionality for the Multimodal Search are highlighted in orange and Genetic Mixing features are highlighted in turquoise.}
  \Description{A screenshot of depicting the interface for the multimodal search and genetic mixing features of SynthScribe.}
  \label{fig:preset_search}
\end{figure*}

\subsection{Multimodal Search}
\label{subsection:preset_retrieval}
With SynthScribe, users can use a flexible natural language query or the existing synthesizer sounds to search through presets (Fig.~\ref{fig:preset_search}). This method is inherently agile, allowing users to rely on the first timbral descriptors that they think of and then refine their search by selecting a synthesizer sound that most closely matches their intentions. Our approach returns the results as a list of presets that automatically configure the parameters of the synthesizer when clicked. Users can search through the current preset bank using text by typing a query into the text box and clicking \textit{"Search Presets"}. In general, we recommend that users type queries using the form "the sound of a ..." due to the training data of the deep learning model in the backend \footnote{For a detailed explanation on the format of the training data, see section 2.3 in \cite{laionclap2023}}.  Users can then refine this query by clicking on a preset in the list and clicking \textit{"Audio Search"}. The audio search feature will return presets which are closest to the current selected preset (the given preset will appear at the top of the list and the list of presets is turned turquoise to indicate that an audio search has been executed). The intention of this feature is to allow users to iteratively get closer to a relevant result by starting with general results from a text query and then narrowing the search space by having them run audio searches on sounds that are closer to their intended sound.

\subsection{Genetic Mixing}
By double-clicking on a retrieved preset, users can collect their favourite sounds in the \textit{"Favourites"} list. If desired, users can opt to create hundreds of new synthesizer sounds by mixing and matching the parameters of these sounds using the Genetic Mixing Algorithm (Fig.~\ref{fig:preset_search}). This is commenced by clicking the "Mix" button. The mixing process scales quadratically with the number of input presets and takes approximately a minute to mix 5 presets which results in 100 new sounds. Upon completion, the user is presented with a new generation of preset sounds which are named numerically. The user is free to repeat this process any number of times and can navigate through the various generations that they create using the \textit{"Next"}, \textit{"Prev"}, and \textit{"Clear"} buttons where the clear button removes all but the first generation which is the default bank of preset sounds.  For any generation, users can still complete text and audio searches as before, enabling them to rapidly find desirable new sounds within the current generation.

\subsection{Preset Modification}

\begin{figure*}[t]
  \centering
  \includegraphics[width=0.75\textwidth]{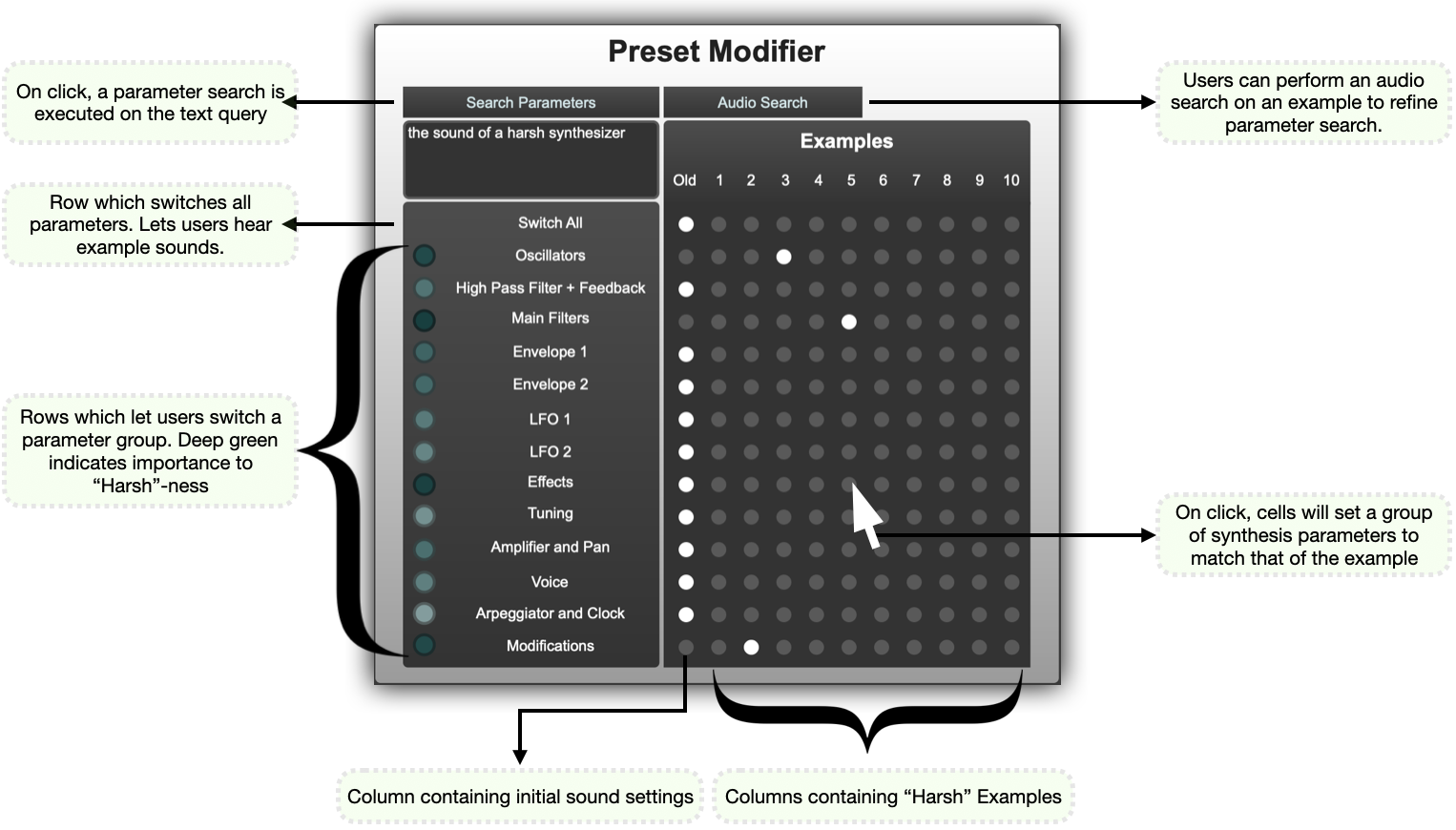}
  \caption{Interface for the Preset Modification feature. Users can retrieve examples that are relevant to a desired effect by first using a text query using the same recommended format as the multimodal search. Examples relating to this text query are placed in columns of the "Examples" matrix. These examples are labelled numerically and can be refined using an audio search. The "old" column contains the sound the user wishes to modify. They can change this sound by clicking on cells in other columns. Each row corresponds to 10 possible changes for a group of parameters and the importance of these groups is highlighted using the green LEDs to the left of the parameter group names.}
  \Description{A screenshot of depicting the interface for the preset modification feature of SynthScribe.}
  \label{fig:modifier}
\end{figure*}

Once a user has found a sound, they can opt to modify it with the preset modification feature (Fig.~\ref{fig:modifier}). This feature contains three key components, the text-search box, the parameter group highlighter (LEDs in various shades of green), and the example matrix. The preset modification process starts with the user typing in a query describing what they are looking to add to their current sound using the same format of text query recommended in section \ref{subsection:preset_retrieval} (e.g. "the sound of an echo"). Upon clicking the \textit{"Search Parameters"} button, the shade of the parameter group highlighters will deepen or lighten. The depth of color determines the importance of that group of parameters with respect to a text query, which helps users determine which parameters should be modified to achieve a desired effect. This directs users to the important rows of the example matrix, where users may click on cells within a row to modify their original sound. Each row is a radiogroup that is responsible for configuring a group of parameters that control a specific aspect of a sound. Each numbered column corresponds to an example preset which  relates to their desired effect while the first column labeled "old" contains the users' initial sound. Users can hear the example sounds in their entirety by clicking on the cells in the first row which changes all synthesis parameters. If the user finds a particular column which certainly contains an example with their desired effect, they can reset the example matrix with an audio search for that column (cells of the matrix are turned turquoise to indicate an audio search has been executed). This audio search will update the parameter group highlighters and  help to highlight important groups by letting the user get specific with their desires.

\subsection{Supporting Features}
\begin{figure*}[h]
  \centering
  \includegraphics[width=\textwidth]{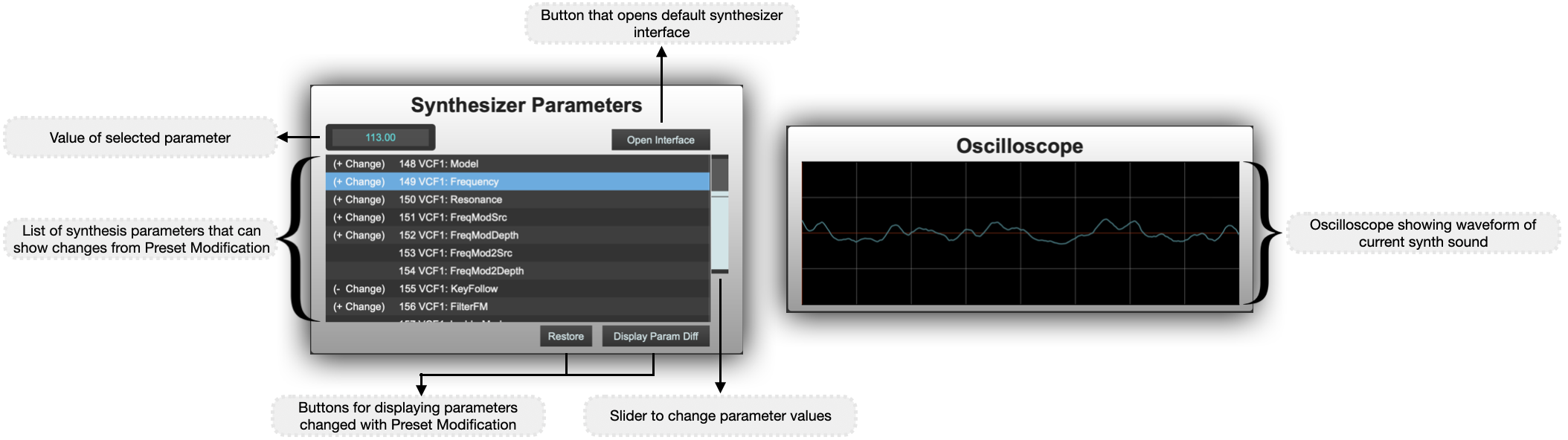}
  \caption{On the left is the Synthesizer parameter display which lets users access low-level synthesis parameters and the default Diva synthesizer interface. On the right is an Oscilloscope which provides a visualization for the shape of the waveform being emitted from the synthesizer.}
  \Description{A screenshot depicting the Oscilloscope and synthesizer parameter display.}
  \label{fig:supporting}
\end{figure*}

We implement two additional features that complement the other available aspects of SynthScribe (Fig. \ref{fig:supporting}). The first is the Oscilloscope which is used to provide a visual reference for the shape of the waveform that the synthesizer is generating. This can provide users with an intuition for how the Preset Modification feature is changing their sound. Additionally, we allow the users to scroll through and modify individual parameters in the Synthesizer Parameters panel. When clicking on a parameter in the list, the users are presented with a slider or a set of radio buttons that allow them to modify continuous or discrete parameters, respectively. Users can also limit the parameters in this list to only the parameter groups that have been changed by the Preset Modification feature. This affords the opportunity for further exploration of synthesis parameters which have been directed by changes made using Preset Modification.

\section{Backend Implementation}
We built SynthScribe using a combination of Max \cite{max} --- a visual programming language used by musicians to develop bespoke synthesis algorithms with interface development capabilities --- and Python. Max is used to host the Diva Synthesizer and to handle notes played on digital instruments which are sent to the Diva Synthesizer to be rendered as an audio signal. Python is used to implement features which require machine learning capabilities and to provide support when a programming task is difficult to implement in a language like Max. These Python functions are executed with inputs from Max by making POST requests to a Flask~\cite{grinberg2018flask} API. All of our ML-related features make use of LAION-CLAP embeddings to connect text and audio. These embeddings represent the modalities of text and audio in a joint space, thus allowing for the retrieval of audio with text while also allowing for direct comparisons to be made between audio samples. We choose a checkpoint \footnote{$music\_audioset\_epoch\_15\_esc\_90.14.pt$ available at \url{https://github.com/LAION-AI/CLAP}} which has been trained first on general audio and then finetuned on music, allowing users to describe instruments in addition to other non-musical sounds when searching for presets. The system is implemented in its entirety on a 2021 Macbook Pro with an M1 Max chip. Below we describe the functionality of the key components.

\subsection{The Diva Synthesizer}
We purchased the Diva Synthesizer by U-he to create a foundation for our system. We chose this synthesizer as it is relatively popular and has been used for similar tasks in a relevant prior work \cite{esling_flowsynth_2020, esling_universal_2019}. In addition to this, it also has a large amount of preset sounds which are available for free on the internet. In total, we make use of the 1200 factory preset sounds made by U-he and downloaded 2328 user preset sounds created by independent musicians on the internet.

\subsection{Multimodal Search}
\label{subsection:preset_retrieval_backend}
To enable text and audio-based preset retrieval, we make use of LAION-CLAP embeddings. For the existing presets, we record and embed them ahead of time using a combination of Max and Python. Each preset is recorded at middle C for 4 seconds with the note being sustained for the first second of that interval. We choose middle C due to its location at the centre of typical frequencies used in music and have the note sustained for a second to get a sense of how the amplitude of the sound changes when a note is held and then released. For a text search, a Flask API endpoint handles a POST request from Max containing the text query. This text query is then embedded and the embedded presets are ranked with respect to their cosine similarity to the text embedding. This ranked list is then returned to Max to reconfigure the ordering of the preset list. If the user wants to execute an audio search, the embedding for the preset they have selected is retrieved and the presets most similar to the given preset (including the given preset itself) are returned to the user.  

\subsection{Genetic Mixing}
\begin{figure}[H]
  \centering
  \includegraphics[width=0.9\linewidth]{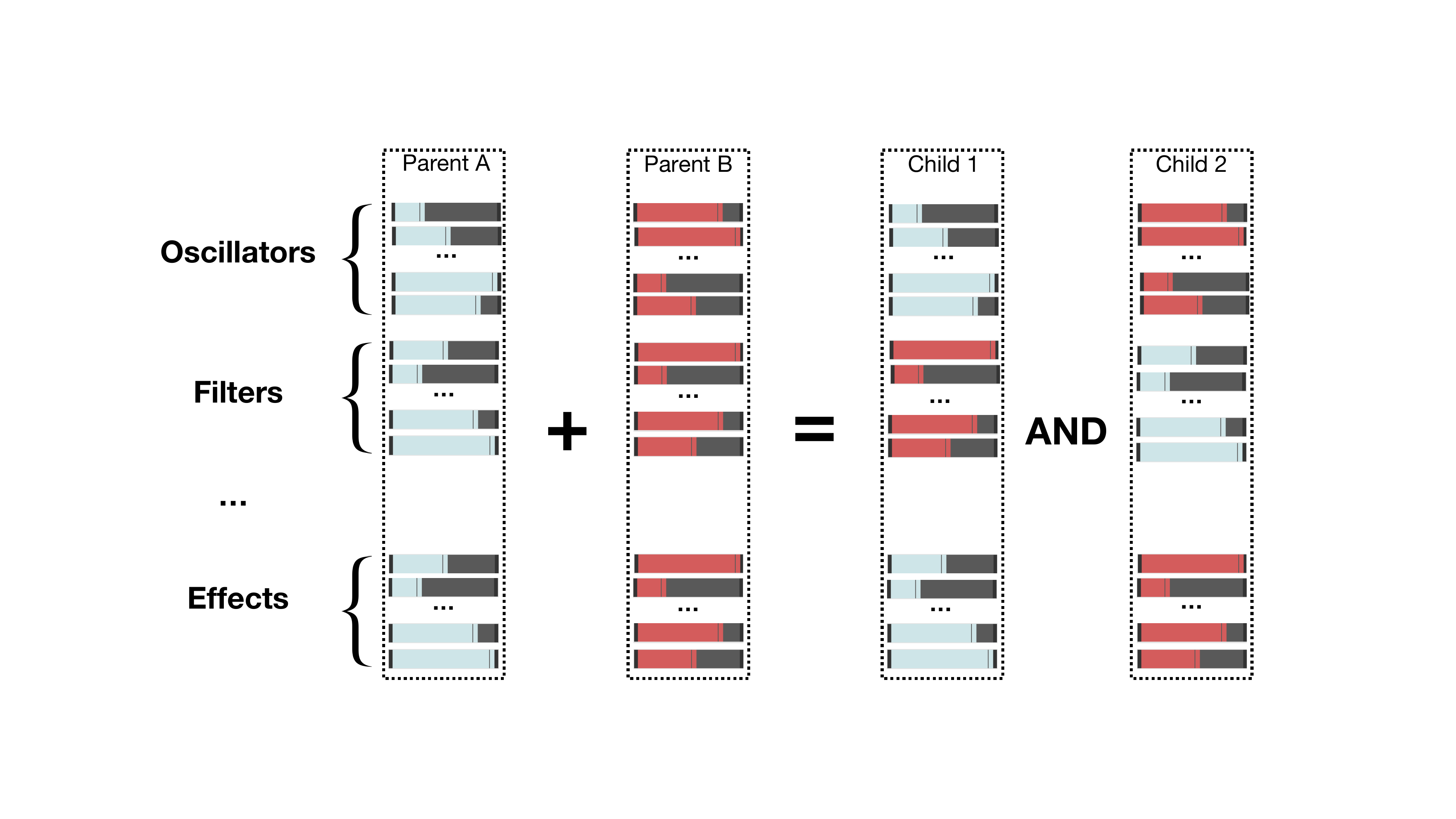}
  \caption{Demonstration of a uniform crossover between two parent synthesizer sounds. The groups of parameters (Oscillators, Filters, ..., Effects) are swapped in their entirety to create the children.}
  \Description{A visualization that demonstrates the uniform crossover algorithm used for the genetic mixing feature.}
  \label{fig:crossover}
\end{figure}

The genetic mixing feature can be described using the language of Genetic Algorithms (GA). Initially, the user selects a group of fit presets to be bred together in order to create a new generation of fit presets. To create this new generation of individuals, we breed pairs of preset sounds by having a child randomly inherit whole groups of parameters from one parent or another (Fig.~\ref{fig:crossover}). The parameters of Diva and many other synthesizers are broken down into groups of parameters that control specific aspects of the sound. For Diva, we recognize 13 such groups. In a breeding operation between parent A and parent B, two children are always created. For a specific group of parameters, the first child will randomly inherit either parent As or parent Bs group with equal probability. The second child will always receive each group that the first child did not inherit. For each pair of parent presets, we complete 5 breeding operations resulting in a total of 10 children per pair. This operation is equivalent to a uniform crossover over the groups of parameters. Upon creating a new generation of presets, they must be recorded and embedded to maintain search functionalities. This is achieved by first using the Python library DawDreamer~\cite{braun2021dawdreamer} to record each preset sound faster than real-time and then using LAION-CLAP to embed each of these new recordings.

\subsection{Preset Modification}
The preset modification feature helps users modify an existing sound to achieve a desirable effect. The interface allows users to search for and discover useful examples that have that effect and then highlights which parameter group from that example should be used to impress that effect on the current sound. The backend makes the relevant examples available and sets the shades of green for the parameter group highlighter. 

\subsubsection{Example Retrieval}
Examples are retrieved either by text queries entered in the text box or by running an audio search on an example contained in a column of the example matrix. The examples are retrieved using the same multimodal search functionality as before. Upon executing either search, the parameters of the examples and of the current sound are saved so that they can be referenced during on-click interactions with the example matrix. Upon finishing the search, the top 10 most relecant results are made available in the numbered columns and the original sound is accessible under the "old" column.

\begin{figure}[H]
  \centering
  \includegraphics[width=\linewidth]{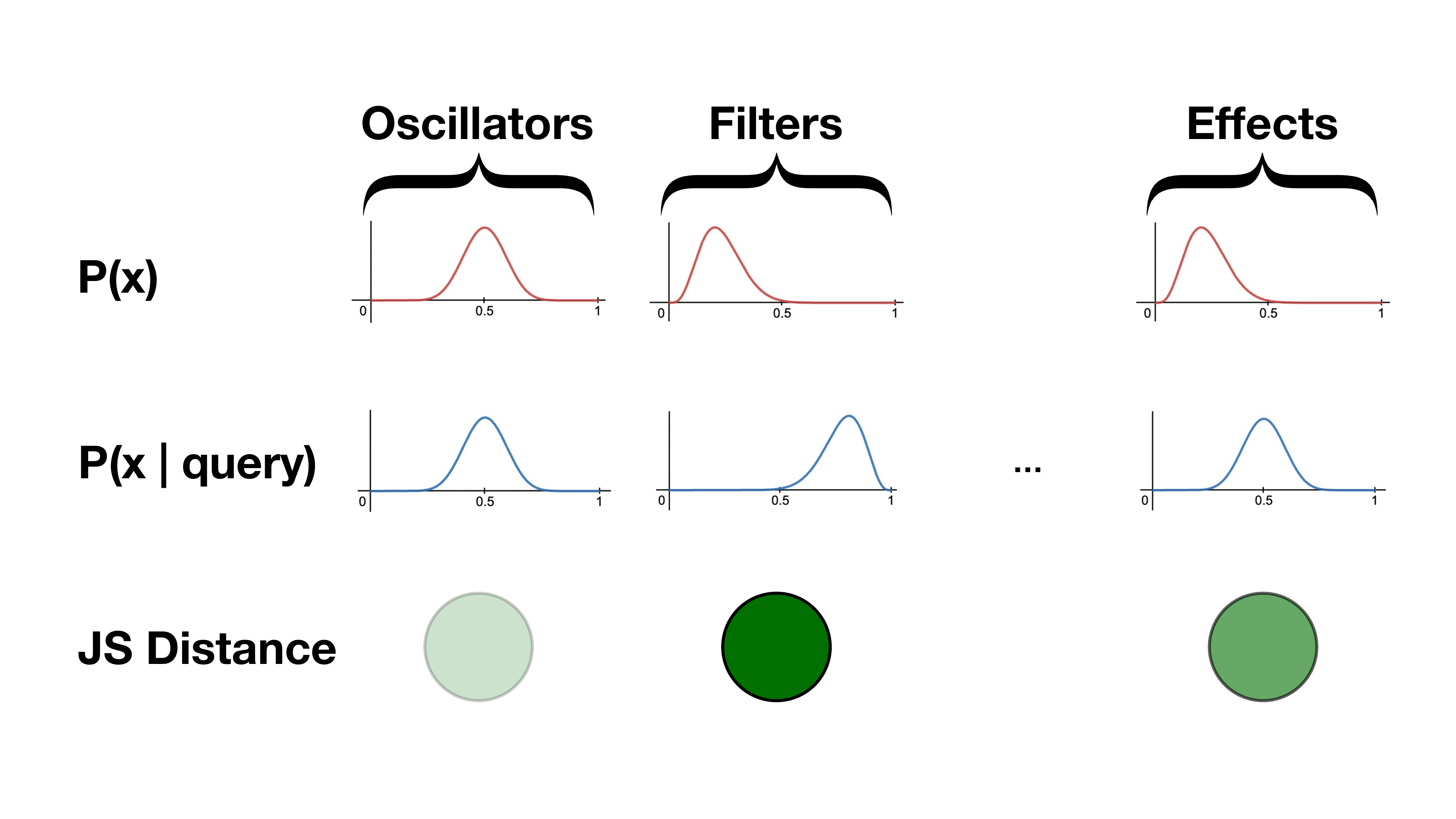}
  \caption{A visualization to show how the importance of each parameter group is calculated. This shows how the shade of green is deepest when the distribution for the parameter is greatly changed as we isolate the synthesizer sounds returned with respect to some query.}
  \Description{A visualization explaining how parameter importance is calculated.}
  \label{fig:js_dist}
\end{figure}
\subsubsection{Example Matrix Interactions}
On-click interactions with the example matrix result in several parameters on the Diva synthesizer being adjusted simultaneously. When clicks occur in the first row, the backend sends a string to Max which changes all of the parameters in the synthesizer to match the selected column in the example matrix. When clicks occur outside of the first row, the backend returns a string which changes the parameters of that specific parameter group to match those of the example selected. Any clicks in the "old" column always result in the restoration of some of the parameters of the original sound with all of the parameters being restored when that click occurs in the first row.

\subsubsection{Parameter Group Highlighter Calculation}
The parameter group highlighter indicates the importance of a group of parameters with respect to a query. It provides the foundation for having the user modify presets because it guides the users' exploration of the example matrix. We implement this by treating each parameter on the synthesizer as a random variable and observing differences in the distributions of a parameter over the whole preset bank against an approximate distribution of that parameter for the top 100 presets returned for the user's query (Fig. \ref{fig:js_dist}). If there is a large difference in these distributions, it shows that this parameter is in part responsible for the perceptual qualities in the sounds returned for this query. For example, if the user searches for "the sound of an arpeggio" it can be expected that the parameters in the top 100 presets for the Arpeggiator will more often be turned on and modified in a unique configuration than those returned outside of the top 100. To quantify this difference, we first approximate the distributions of each parameter on the synthesizer using all 3528 presets. Discrete parameters are left as is while continuous parameters (always valued between 0 and 1) are approximated with 10 equidistant regions or bins. Empty bins are handled with additive smoothing. Most continuous variables have a default value which are large spikes in probability density at the extremes or precisely in the middle of a parameter range. To better represent this reality, we give default values their own narrow bin. At inference time, we take in a text or audio query from the user and find the top 100 presets using the cosine similarity of LAION-CLAP embeddings. We then recalculate the distributions for each parameter over these 100 samples using the same bins and additive smoothing as before. These distributions are then compared using Jensen-Shannon Distance \cite{jensen_shannon_distance} --- a stable and symmetric distance measure between distributions which ranges between 0 and 1. We proceed to calculate the average distance for the top 20 parameters with the largest distance in each group. We select the top 20 so that the importance of large parameter groups is not diluted by the fact that many of their parameters will be left at default settings. Upon calculating this average for each group, we assign the parameter group with the largest average distance to the deepest shade of green and interpolate the colors for other groups on a range between 0 and the maximum. We choose this colour interpolation strategy to ensure that the maximum is always obvious.

\section{User Evaluation}

\label{section:quant}
We evaluated our multimodal search and preset modification features using two tasks and one participant group by soliciting subjective ratings that allowed us to quantify their performance. For the multimodal search, we evaluated LAION-CLAP's ability to retrieve synthesizer sounds and used BERT~\cite{devlin-etal-2019-bert} as a strong baseline. For the preset modification features, we asked participants to listen to modified sounds suggested by our system which allows us to evaluate the quality of the modifications and the effectiveness of the parameter group highlighter simultaneously.

\subsection{Participants}
\label{subsection:participants}
We recruited 8 participants in total for the study (Mean age=25.0, STD=2.9). Participants were not required to have any level of musical experience. Only two participants actively played an instrument and none had extensive experience using synthesizers. In this regard, we followed prior work which used inexperienced participants for discerning differences in timbre. \cite{huang2018timbretron}. We leave engagement with musicians for our free usage observations in section \ref{section:free_usage}. The study took place across two independent tasks which happened on different days where each task lasted approximately 30 minutes. Participants were compensated with a total of 20 CAD.

\subsection{Methodology}
\subsubsection{Task 1: Preset Retrieval Evaluation}
Participants were tasked with rating the relevance of a set of synthesizer sounds with respect to a text query on a 7-point Likert scale. In total, each participant was given two text queries and evaluated a total of 10 sounds for each text query. The text queries were generated before each study in a quasi-random fashion using an adjective and an instrument class to describe a sound. An example query could be "The sound of a harsh brass instrument". The list of possible adjectives was compiled from relevant research in music psychology and keywords in text labels of synthesizer sounds from commercial sources. The list of possible instrument classes was sourced only from keywords used in keyword searches for synthesizers (Both lists are available in~\ref{subsection:instruments}). To make these queries understandable to our participant class, we avoided instrument classes and adjectives that would be non-intuitive to people without musical experience. For each query, we retrieved 5 sounds using LAION-CLAP and 5 sounds using BERT from the Diva Synthesizer preset bank. LAION-CLAP retrieval was achieved using the strategy described in subsection~\ref{subsection:preset_retrieval_backend}. BERT retrieval was implemented by extracting adjectives and instrument classes for each sound in the Diva Synthesizer preset bank using GUI automation and OCR. Given these adjectives and instrument classes, we embedded sentences of the form "The sound of an {adjective} {instrument class}" using BERT. When querying both models, we used the form "The sound of..." to better fit with the training data for LAION-CLAP; however, since we used it to create out BERT embeddings as well this gave LAION-CLAP no undue advantage for the task. The five sounds were then compiled into a list of 10 in a random order. The participant was then allowed to play each sound on a musical keyboard and then provided their subjective evaluation of the sound with respect to a text query.

\subsubsection{Task 2: Preset Modification Evaluation}
\label{subsubsection:task_2}
Participants were tasked with rating the quality of a modification of a synthesizer sound with respect to an adjective. This process started by having the user choose an initial sound from the Diva Synthesizer preset bank which resembled one of the instrument classes in \ref{subsection:instruments}. The participant was then given their adjective (e.g. harsh) which was searched for using the text input in the preset modification feature. Searches were executed using the form "the sound of an {adjective} synthesizer" to better represent the training data of LAION-CLAP. The user then listened to all of the example columns from the preset modification feature and selected the column that best fit with their adjective (e.g. the harshest sound from the examples). An audio search was then ran on that example which reconfigured the examples to be the sounds most similar to their chosen column and also put their chosen column in the "1" position. Participants then compared their original sound with modifications of that sound for all thirteen parameter groups, rating the quality of a modification on a 7-point Likert scale. If their adjective was harsh, a rating of 1 suggested that the modification hadn't changed anything or had made the sound less harsh while a rating of 7 corresponded to a modification that made the sound notably more harsh. The user repeated this process twice for a total of 26 ratings of modified sounds.

\subsection{Results}
\subsubsection{Results for Task 1}
As shown in Table \ref{tab:task1}, LAION-CLAP outperformed BERT with a higher average and median rating. Further analysis with a Wilcoxon Signed-Rank test showed that this result is statistically significant ($Z = -2.392$, $p < .05$).

\begin{table}[h]
    \centering
    \includegraphics[width=.475\textwidth]{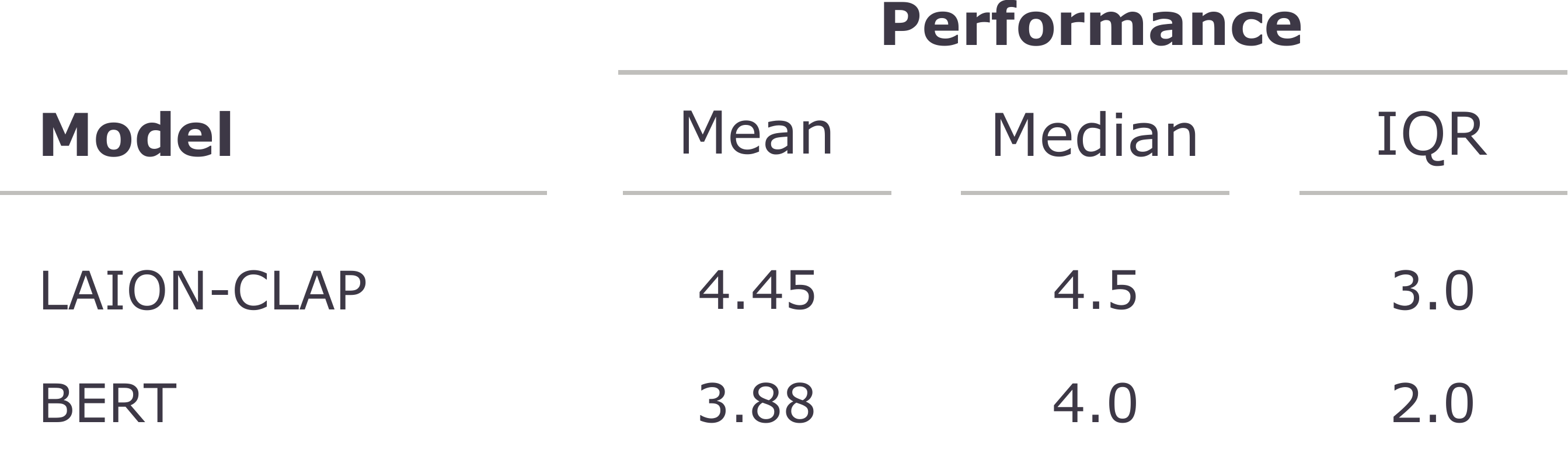}
    \caption{The results of the comparative evaluation of LAION-CLAP and BERT show that LAION-CLAP outperforms BERT. The results of a Wilcoxon Signed Rank test show that these results are statistically relevant ($Z = -2.392$, $p < .05$). }
    \label{tab:task1}
\end{table}

\subsubsection{Results for Task 2} The median value of the maximum score for set of modifications created using the procedure from \ref{subsubsection:task_2} is quite high (Median 6.0, Average 6.0, IQR 0.5). Therefore, the best modifications suggested by SynthScribe are typically relevant to a given adjective. We also evaluated the parameter group highlighter by comparing populations of ratings for possible group modifications that occurred in the top 5 group modifications recommended by the highlighter feature and those outside of the top 5. The top 5 modifications were rated better than the lower-rated modifications with this difference being statistically significant ($U = 6419.0$, $p < .05$) after performing a Mann-Whitney U test. Both ratings are quite low due to the fact that some of the modifications have little to no effect. This shows that the parameter group highlighter feature provides some benefit but cannot predict which group of parameters will give the best modification reliably.

\begin{table}[h]
    \centering
    \includegraphics[width=.5\textwidth]{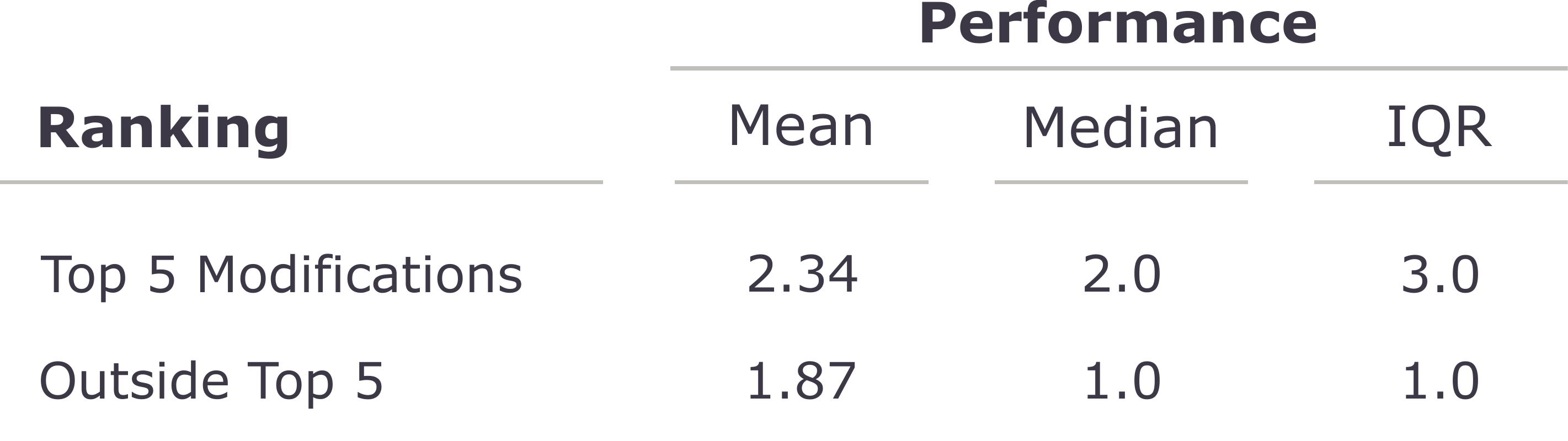}
    \caption{This table presents the analysis of the Parameter Group Highlighter, showing a difference in ratings for modifications in the top 5 and those outside of the top 5. The results of a Mann-Whitney U test show that the difference between these groups of ratings is statistically significant ($U = 6419.0$, $p < .05$)}
    \label{tab:task2}
\end{table}

\section{Free Usage Observation with Musicians}
\label{section:free_usage}

\subsection{Participants}
Our study consisted of 6 musicians who had experience creating music. Two were professionals that make music for a living (P1, and P2) and the rest were hobbyists who made music for their own enjoyment.  The professionals had participated in the formative studies and there was no overlap with the participant group in subsection \ref{subsection:participants}. Both of the professional musicians had received formal instruction on synthesizers in an academic setting with P2 describing themselves as an expert and P1 an intermediate. The hobbyists had at most encountered synthesizers when making music and described themselves as either novices or beginners. All musicians had at least basic keyboard skills. Each were compensated with 20 CAD.

\subsection{Methodology}
The study consisted of a 15-minute demo of SynthScribe followed by the participant completing at least two independent musical tasks. The participant was provided with a 61-key keyboard that they used to play the synthesizer sounds. P2 brought in and was permitted to use their Akai EWI 5000~\footnote{Akai EWI 5000: \url{https://www.akaipro.com/ewi5000}} --- a digital wind instrument. The participant was free to select any task that they might encounter while using a synthesizer for their own purposes. Suggested tasks included finding a desirable synthesizer sound for a song they knew how to play, replicating a synthesizer sound from a song, or finding a synthesizer sound that might fit in a score for a film of a particular genre. When using the system, we requested that the participant write text queries in the form of "The sound of a ..." to ensure the best results due to the nature of the training data for LAION-CLAP. The study concluded with a 15-minute semi-structured interview followed by a short survey in which we solicited subjective ratings for ease of use and usefulness in addition to collecting NASA-TLX~\cite{hart1988development} ratings for frustration, effort, and mental demand.

\subsection{Results}
\begin{figure*}[t]
  \centering
  \includegraphics[width=\textwidth]{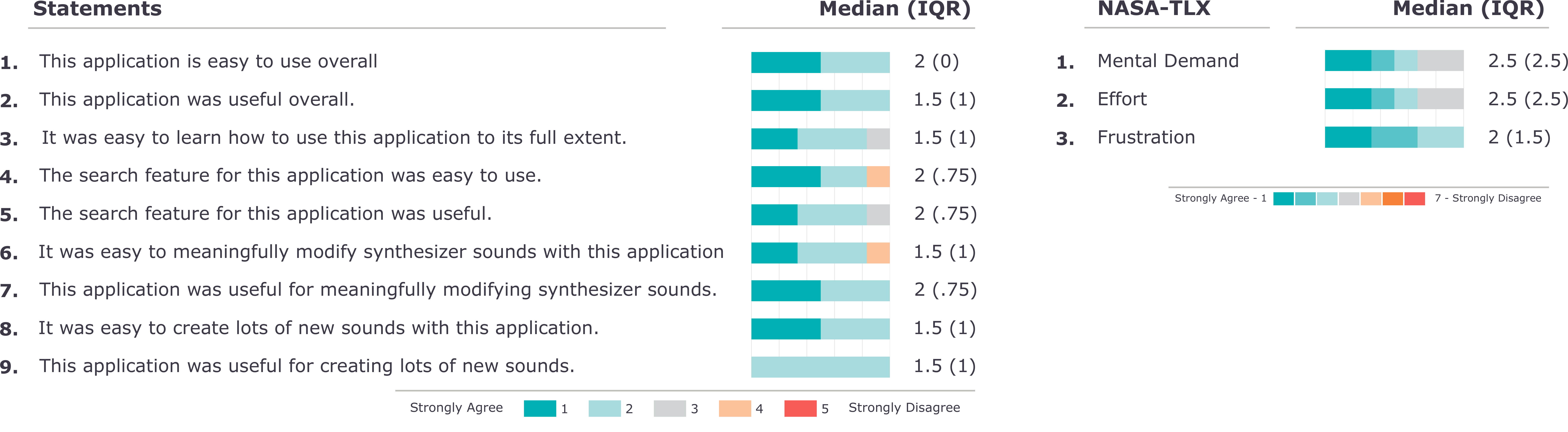}
  \caption{Results from the questionnaire, and NASA-TLX ratings (Median and Inter-quartile Range). Note that the first two were rated with a 5-point Likert scale and NASA-TLX was completed on a 7-point Likert scale.}
  \Description{This displays the results of the questionnaire. Participants widely agree that each feature is useful or easy to use with one participant disagreeing that the search feature was easy to use and another disagreeing that it was easy to meaningfully modify the synthesizer sounds with SynthScribe.}
  \label{fig:feature_ratings}
\end{figure*}

In Fig.~\ref{fig:feature_ratings} we summarize the subjective ratings of SynthScribe. Our participants broadly agreed that SynthScribe is easy to use and useful overall in addition to imparting relatively low mental demand and effort while causing little frustration. The overall results of these studies were positive. P2, a professional musician, mentioned that "even though like I know a lot about synthesis and can probably get pretty far with just the DIVA interface, this actually saved me a lot of time." The other professional musician, P1, mentioned that the system would be most useful for them when learning to use a new synthesizer. Participants (P2, P3, P5) also mentioned that they enjoy the discoverability aspects of the system and described situations where they arrived at fitting but surprising new sounds. We also received invaluable constructive feedback that can inspire future directions. Below, we further analyze the feedback we received on each feature.

\subsection{Multimodal Search}
We observed that some participants use the multimodal search in a step-wise fashion (P3, P5, P6). First, they ran a text search with a general description of their idea and then they ran an audio search on a sound they like. P5, for example, started by searching for the sound of a piccolo and then narrowed their search by running an audio search on a sound they liked which led to a recorder sound that they felt to be most useful. P3 described the advantage of a general initial text search when they were looking for ambient sounds: "My goal was very loose and it gave me some things that, for example, the bass being a candidate for the kind of ambient sound I want, I wouldn't have thought of that myself." Participants sometimes had difficulties with the query format. P4, for example, thought typing "the sound of ..." every time was "completely irrelevant." They also suggested that it might be better to have a pop-up window containing the results of an audio search to avoid losing the original content of the initial text search.

\subsection{Genetic Mixing}
The Genetic Mixing feature was appreciated by participants for the pleasant but unexpected sounds it created (P2, P3, P4, P5). P4 elaborated on this by saying "I think the mix feature allows you to kind of have these weird overlaps that you wouldn't normally have in a synth and that was really cool." The fact that our mixing algorithm achieved aesthetically pleasing and surprising results is bolstered by the reviews of the Multimodal Search due to the fact that users can search through new generations of sounds using the same search strategies as before. Some participants expected the mixes to be slightly more thorough. P2, for example, noted that they expected the Genetic Mixing to do some level of interpolation between the continuous parameters in the favorites list instead of just swapping parameters. P3 mentioned that a useful addition would be a system that automatically names the new synthesizer sounds as opposed to the current generic naming convention.

\subsection{Preset Modification}
The preset modification feature was used by participants to make quick directed changes. Both professional musicians, P1 and P2, used this feature to make the final adjustments when attempting to replicate synthesizer sounds from songs they'd chosen. To this end, P2 claimed that "even though like I know a lot about synthesis and can probably get pretty far with just the DIVA interface, this actually saved me a lot of time. P5 noted that when using synthesizers in the past "the search space seemed unlimited" but that this feature provides "a really fast way to just iterate through all of the different possibilities out there". They also claimed that the parameter group highlighter was a useful indicator of importance. Other participants highlight some non-intuitive aspects of this feature. P4 and P6 believed the numerical labeling of the examples corresponded to an increasing quantity. P3 felt that having to use nouns in the text queries (e.g. "the sound of a harsh synthesizer") required more mental effort than just typing an adjective or command (e.g. "more harsh"). P4 outright stated that they disliked the feature because it felt like it required them to understand the functionality of the different parameter groups.

\section{Discussion and Future Work}
\subsection{Facilitating Synthesizer Use with Multimodal Deep Learning}
Our quantitative results from section \ref{section:quant} show that LAION-CLAP is capable of reliably returning relevant synthesizer sounds for the queries we tested. Although this model is not designed explicitly to work with synth sounds, these results show that SynthScribe is built on a solid foundation, allowing researchers to take inspiration from our approaches when more suitable multimodal models become available. These results also show that users should be able to find a relevant modification of their sound while using the preset modification feature due to the fact that the best modifications for a query are rated quite highly. The parameter group highlighter, however, does not always direct users to the most useful modifications. In developing this application, we tested the group highlighter for situations in which it could create an objectively correct result. For example, if we want to give the sound an echo searching for "echo" should highlight the "Effects" group as most important. It worked well for these use cases but we believe this is because the "Effects" parameter group is either on or off and there are lots of examples of synthesizer sounds with an echo. Over the top 100 samples returned for the query "echo", it's likely that the majority of examples would have the echo turned on, leading to a stark difference between the top 100 synthesizer sounds and the overall distribution of synthesizer sounds. In situations where a user is looking for a modification for which there are only a few good examples, it is likely that lots of the synthesizer sounds returned in the top 100 will not be relevant to their query, leading to the importance highlighter being diluted with irrelevant examples. 

Our free usage observations provide examples of how and when these features can be useful to musicians of various experience levels. The multimodal search is shown to provide a useful paradigm that allows musicians to describe their desires at a high level in text and then specify their desires further by using an audio search. With the Genetic Mixing feature, we observe promising interactions that provide strong anecdotal evidence that new exciting sounds can be created by mixing those from a favourites list. Users emphasize their joy in finding something surprising, an observation that we describe in more detail in the next section. Finally, the preset modification figure is shown to help people make relevant changes to their sounds with some musicians emphasizing the time they could save when using this feature. Underlying these features, we show that a foundation of multimodal deep learning is useful in the development of such a system. Given the efficacy of our features, we are able to develop an interface around an alternative control modality for synthesizers without having to train a new model or collect user annotations on the fly.

Below we list some suggestions for future work that may address some limitations of SynthScribe. SynthScribe is built on top of one synthesizer but in practice, musicians interact with many synthesizers. It would be useful to design a system with similar features that manages many synthesizers simultaneously. It could also be useful to tailor a multimodal deep-learning model to synthesizer sounds. This could include training a model on text-audio pairs including synthesizer jargon or common words that people use to describe timbres (e.g. dark, bright, mellow, harsh, etc.). It's also important to note that people tend to describe sounds with their own idiosyncrasies which could lend this design problem to systems that leverage personalization. To highlight which parameters are important to make a desirable modification, it could be useful to collect a larger bank of synthesizer sounds which would prevent the failure case that we describe above. It could also be useful to calculate importance based on fewer examples which would prevent unrelated examples from clouding the calculation. Additionally, relevant prior work has already focused on developing intuitive sliders for synthesizers~\cite{huang_active_2014} but used a synthesizer with relatively few parameters and required user annotations. Future work might use the preset modification approach to help the user isolate a subset of important parameters quickly which could then be used in tandem with other techniques to provide intuitive sliders with fewer user annotations and effort.

\subsection{Emphasizing Surprise}
In our free usage observations, several participants emphasized that SynthScribe features often returned surprising but relevant results. Whether it was a sound retrieved with the multimodal search or encountered via Genetic Mixing or Preset Modification, participants emphasized their enjoyment of unexpected but pleasant sounds. Connecting this with our formative study, this could be due to the fact that finding an unexpected sound is correlated with the pursuit of originality when creating new music. This result is also affected by the context of the user study in which a musician is freely working with a synthesizer. We speculate that it may also be that musicians want synthesizer sounds to surprise them, sparking new ideas and creative directions. 

To imagine how future work might incorporate this, we would like to highlight a fictional counter-example. SynthGPT ~\footnote{SynthGPT: \url{https://www.youtube.com/watch?v=wcFmbjQ_mEc}} is a fake system which had a spoof demo video posted on YouTube. SynthGPT functions by providing musicians with "any sound imaginable simply by typing a text description". Given our observations, we hypothesize that a system like this one would leave much to be desired. In this creative context, it seems that helping users find what is imaginable is just as important as helping them discover the unimaginable. In SynthScribe, we do this with the genetic mixing feature by getting unexpected but desirable sounds by mixing sounds that musicians have labeled as their favorites. A more refined approach might attempt to learn what is surprising to a musician and also what they like, allowing for unimaginable and desirable sounds to be discovered automatically.

\section{Conclusion}
We implemented a novel system, SynthScribe, which used multimodal deep learning as a foundation for several features designed to facilitate the use of synthesizers for musicians. The results of our user studies show that our system is capable of helping musicians search for and modify synthesizer sounds effectively while also affording them the opportunity to create new and surprising sounds that inspire creativity. We outline that future work can aim to tailor multimodal models to the language used to describe timbres but may also benefit from incorporating aspects of personalization. We highlight that our participants particularly enjoyed sounds that they liked but did not expect. We recommend that future work may also try new methods that predict sounds that musicians will enjoy but were not explicitly asked for.
\section{Acknowledgements}
We would like to acknowledge the guidance we received from various collaborators that shaped the direction of SynthScribe. Particularly Nicolas Jonason who referred us to LAION-CLAP as a candidate foundation model and Marcel Castillo who provided a breadth of musical perspective in the early stages of the project. We would also like to thank Dani Oore who remixed one of his and Andrew Staniland's compositions by using SynthScribe to source new preset sounds \footnote{Mythos Synthscribe Remix: Deus In Machina by Dani Oore and Andrew Staniland: \url{https://soundcloud.com/dani-oore/mythos-synthscribe-remix-deus-in-machina}}.

\bibliographystyle{ACM-Reference-Format}
\bibliography{references}

\newpage
\appendix
\section{Appendix}

\subsection{Quantitative User Study Instruments and Adjectives}
\label{subsection:instruments}
The list of possible instruments was determined by examining the classifications of sounds included in Analog Lab V \footnote{\url{https://www.arturia.com/products/software-instruments/analoglab-v/overview}}. The list of possible adjectives was determined from Analog Lab V, the keywords used in the Diva Synthesizer's keyword search, and a research paper on the language used to describe timbre \cite{adjectives}. We narrowed down both lists to examples that we felt would be understood or at least easily explainable to our participant population for this study which was comprised of mostly non-musicians. This meant not using words that included synthesizer-specific jargon in the instruments and ignoring adjectives (bright, dark, e.g.) that we believed would be non-intuitive to our participant population. 

\begin{table}[h]
    \centering
    \begin{tabular}{|p{0.2\linewidth}|p{0.8\linewidth}|}
        \hline
        Instrument Class & Wind Instrument, Electric Piano, Bass, Drum, Brass Instrument, String Instrument, Organ \\
        \hline
        Adjectives  & Percussive, Constant, Moving, Clean, Dirty, Soft, Aggressive, Thin, Complex, Funky, Sharp, Simple, Punchy, Huge, Bizarre, Mellow, Atmospheric, Airy, Evolving, Short, Long, Noisy, Glitchy, Arpeggiated, Distorted, Acoustic, Dull, Loud, Low, Rough, Smooth, Clear, Rich, Nasal, Full, Hard, Weak, Muffled, Resonant, Large, Quiet, Calm, Harsh, Shrill, Powerful, Metallic, Ringing, Deep \\
        \hline
    \end{tabular}
    \label{tab:words}
\end{table}

\subsection{SynthScribe Interface}
\begin{figure*}[h]
  \centering
  \includegraphics[scale=0.5, angle=-90]{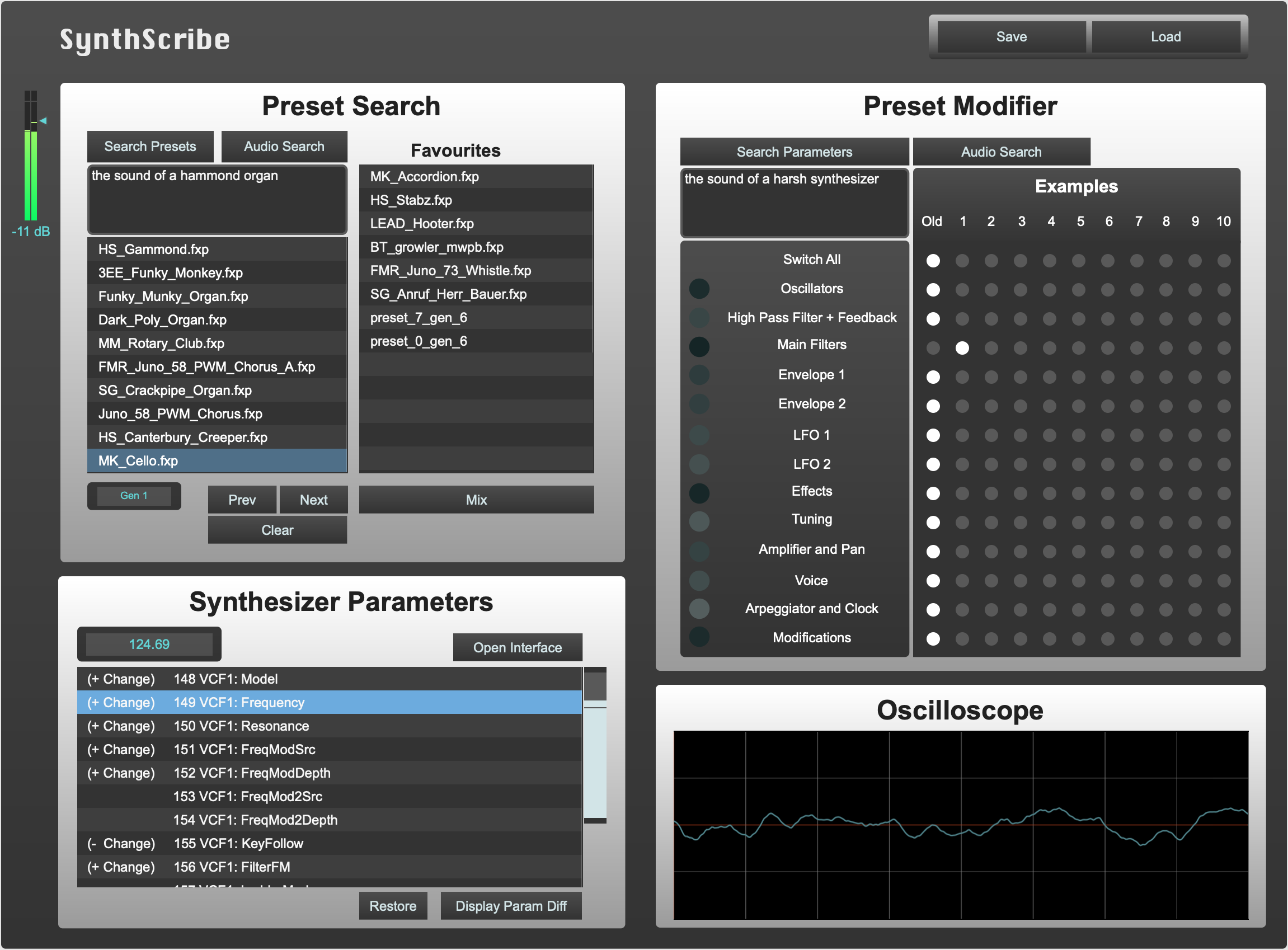} 
  \caption{The whole interface of SynthScribe depicting the layout of the features}
  \Description{}
  \label{fig:interface}
\end{figure*}

The whole interface is depicted on the following page (Fig. \ref{fig:interface}).

\end{document}